# Old Supernova Dust Factory Revealed at the Galactic Center

R. M. Lau[1]*, T. L. Herter[1], M. R. Morris[2], Z. Li[3], J. D. Adams[1,4]

[1]Department of Astronomy, Cornell University, Ithaca, NY 14853, USA

[2]Department of Physics and Astronomy, University of California, Los Angeles, 430 Portola Plaza, Los Angeles, CA 90095, USA

[3]School of Astronomy and Space Science, Nanjing University, 22 Hankou Road, Nanjing, Jiangsu, 210093, China

[4]SOFIA Science Center, Universities Space Research Association, NASA Ames Research Center, MS 232, Moffett Field, CA 94035, USA

*Correspondence to: ryanl@astro.cornell.edu

**Dust formation in supernova ejecta is currently the leading candidate to explain the large quantities of dust observed in the distant, early Universe. However, it is unclear whether the ejecta-formed dust can survive the hot interior of the supernova remnant (SNR). We present infrared observations of ~0.02 $M_\odot$ of warm (~100 K) dust seen near the center of the ~10,000 yr-old Sgr A East SNR at the Galactic center. Our findings signify the detection of dust within an older SNR that is expanding into a relatively dense surrounding medium ($n_e$ ~ 100 cm$^{-3}$) and has survived the passage of the reverse shock. The results suggest that supernovae may indeed be the dominant dust production mechanism in the dense environment of early Universe galaxies.**

**One Sentence Summary:** We present the first evidence of dust within an "old" (~ 10,000 yr) supernova remnant, which supports the paradigm that supernovae are an important dust factory for the earliest forming galaxies in the Universe.

The search for the dominant formation mechanism of large quantities of dust detected in galaxies of the early Universe remains an open investigation of profound importance. It both influences observed emission and greatly affects the formation of future generations of stars. Due to the short lifetimes of their progenitor stars and to their highly metal-enriched ejecta, supernovae (SNe) are believed to be efficient sources of dust production (*1*). However, the powerful SN explosions and the resulting shocks are predicted to also be very effective at destroying and shattering dust: depending on the energy of the explosion and the density of the surrounding medium, less than ~20% of the mass of the SN-condensed dust is expected to survive the passage of the reverse shock that is driven back into the ejecta due to the difference between the thermal pressure of the shocked circumstellar material and that of the expanding ejecta (*2 – 4*). Recent studies have in fact shown that SNe may be net destroyers of dust in present day galaxies, e.g. in the Magellanic Clouds (*5, 6*), but net producers of dust in the earliest forming galaxies in the Universe (*7*). However, there is currently no direct observational evidence of the quantities of SN-condensed dust surviving the passage of the reverse shock through the ejecta.

Sgr A East is the well-studied remnant of a core-collapse supernova, located 8 kpc away near the center of our Galaxy (*8*), that has blown out a ~ 3' (7 pc)-diameter shell of non-thermal radio emission (Fig. 1; *9, 10*). Based on the kinematics of maser spots associated with the SNR, it is estimated to be ~ 5 pc from the Galactic center (*11*). An age of ~ $10^4$ yr has been estimated for the SNR based on predictions of its elongation from tidal forces (*12*) and, more recently, from the observed proper motion and displacement of the neutron star believed to be the remnant of the SN progenitor from the interior of the SNR (*13*). The ~ $10^4$ yr timescale and the metal-rich hot ejecta from the center of the SNR imply that the reverse shock has reached the center of Sgr A East (*13*). Observations of dust associated with the Sgr A East SNR could therefore provide estimates of the fraction of dust that survives the destructive passage of the reverse shock.

We use mid/far-infrared (IR) images of Sgr A East from the Stratospheric Observatory for Infrared Astronomy (SOFIA) to argue the presence of warm ($T_d \sim 100$ K) dust near its center. The warm and dusty environment of the Galactic center (*15*) presents a challenge in definitively demonstrating that the IR emission is indeed associated with the Sgr A East SNR ejecta. Our claim is substantiated by four major results from the observations: 1) The location of the IR emitting region is consistent with the center of the supernova remnant and is spatially anti-correlated with the hard X-ray emission, suggesting that dust is in a cooler region of the ejecta. 2) Analysis of the dust temperatures and heating sources shows that the constraints on its location is consistent with Sgr A East since it must be radiatively heated by the optical and UV photons from the central stellar cluster, which dominates the radiation field in the central parsecs of the Galaxy. 3) The lack of cold dust emission at submillimeter wavelengths coincident with the IR emitting region implies it does not originate from a cloud along the line of sight. 4) We show that the typical size of the emitting grains must be smaller than the dust in the surrounding interstellar medium (ISM), which is consistent with having been processed by the reverse shock of the SN.

We propose that the dust has survived the passage of the reverse shock due to the relatively large density of the surrounding medium, which slows the ejecta, and places it into a temperature and density regime where it will undergo significant radiative cooling on timescales much shorter than dust destruction timescales. With an estimate of the surviving dust mass in Sgr A East, we discuss the viability of SNe as the dominant source of dust production in the early Universe.

**Observations and Results.** The Faint Object Infrared Camera for the SOFIA Telescope (FORCAST) (*16*) was used to obtain images of the Sgr A East SNR at 7.7, 19.7, 25.2, 31.5, and 37.1 μm. The spatially resolved hot and warm dust traced by these wavelength bands is shown in Fig. 2. We utilized *Chandra*/Advanced CCD Imaging Spectrometer imaging array (ACIS-I) to obtain high-quality hard X-ray (2 – 8 keV) images of Sgr A East. Additionally, we incorporated archival mid-IR (5.8 and 8.0 μm) and submillimeter (70 and 160 μm) observations taken by the Spitzer Space Telescope's Infrared Array Camera (IRAC; *17, 18*) and the Herschel Space Observatory's Photodetector Array Camera and Spectrometer (PACS; *19, 20*), respectively (*21*).

**Source Morphology and Location.** The mid-IR (5.8 and 8 μm) and far-IR (19 – 37 μm) observations of the proposed SNR dust emission are co-spatial and confined to the regions near the center of the Sgr A East radio shell (Fig. 1A). The central position of the dust emission is consistent with the expected location of dust having condensed within the ejecta. Furthermore, the dust is spatially anti-correlated with the hard X-ray continuum of the SNR ejecta (Fig 1B). This anti-correlation suggests the dust is located in a much cooler and less hostile region of the ejecta.

It is apparent in Fig. 1A that there is no significant submillimeter emission coincident with the location of the IR emitting region. This is in contrast to the Sgr A East HII regions that are located ~3 pc to the east in projection from the center of the SNR and are the illuminated edges of the molecular cloud associated with the prominent ridge of submillimeter emission that extends along the eastern side of Sgr A East (*22*). The lack of submillimeter emission at the IR emitting region therefore implies it is not associated with a cold, molecular cloud along the line of sight towards the center of Sgr A East.

**Observed Dust Temperature.** A key to understanding whether the dust is located interior to the Sgr A East SNR is its thermal structure. The closest known stellar heating source is the Central cluster of massive young stars surrounding the Galactic black hole, located ~ 3 pc away in projection. We also consider heating via collisions with thermal electrons in the shocked ejecta of the SNR but conclude that it is negligible (*23*). The presence of the strong radiation field from the central cluster therefore presents a unique heating scenario for dust in Sgr A East because the radiative heating contribution, which typically arises from the interstellar radiation field in most SNRs, is usually negligible compared to collisional heating.

In order to investigate the heating source(s) and thermal structure of the possible SNR dust, we generate a color temperature map using the de-convolved 31.5 and 37.1 μm flux maps of Sgr A West and East observed by SOFIA/FORCAST (Fig. 3; *24, 25*). We assume the dust emission is optically thin and takes the form $F_\nu \propto B_\nu(T_d)\,\nu^\beta$, where a value of 2 is adopted for β, which is typical for interstellar dust. The longer wavelength images are used to produce the color temperature map since the signal-to-noise ratio from the IR emitting region at those wavelengths is higher than at 19.7 μm. The average 19/37 color temperature of the IR emitting region is consistent with the average 31/37 color temperature.

It is apparent from the location of the temperature peak and the negative radial temperature gradient centered on Sgr A* that the luminous central cluster dominates the heating of the dust in the HII region and Circumnuclear Disk (CND) immediately surrounding Sgr A* (*24*). Interestingly, the proposed SNR dust exhibits a temperature of ~ 100 K +/- 8 K (*26*), which is much greater than the ~ 75 K that would be predicted for 0.1 μm-sized grains from the radial temperature gradient centered approximately on Sgr A*, allowing for a $\sqrt{2}$ projection factor. We note that the ~ 75 K temperature of the structure we refer to as the Northern Dust Cloud (see Fig. 3) is consistent with this gradient and is equidistant from Sgr A* in projection with the SNR dust. The four Sgr A East HII regions seen to the east of the SNR (Figs. 1 and 3) exhibit temperature maxima, but they are each heated locally by $10^4 - 10^5$ L$_\odot$ sources (*22, 25*).

**Heating Source and Dust Size.** Three different scenarios can be considered to explain why the proposed SNR dust is at a significantly higher temperature than expected from heating by the central stellar cluster: 1) the dust is locally heated by a luminous stellar source or sources, 2) it is heated by energetic electron collisions in the shocked ejecta, and 3) the dust is much smaller in size than the dust associated with the CND and surrounding ISM. The first scenario would probably imply that the dust is associated with an illuminated molecular cloud along the line of sight towards Sgr A East, whereas the latter two scenarios would constrain the location of the dust to the interior of Sgr A East.

First, we consider dominant heating by a local stellar source, as is the case for the Sgr A East HII regions. Given the size of the SNR's IR emitting region (~ 0.8 pc), we assume that any such local heating source(s) would be located at a distance of ~ 0.4 pc from the dust. For 0.1 μm-sized silicate grains that are in thermal equilibrium with the radiation field, the total stellar luminosity of the heating source would then have to be $\sim 2 \times 10^6$ $L_\odot$ in order to reproduce dust temperatures of $\sim 100$ K. A source with such a high luminosity, which is similar to that of luminous blue variables, would be easily detected in the near-IR (1.90 μm; *27, 28*) and near-IR Spitzer/IRAC (*18*) observations of the region; however, no such heating source can be identified within several parsecs of the dust. Additionally, the total integrated infrared luminosity of the IR emitting region is $\sim 7 \times 10^4$ $L_\odot$, which is too high if it were heated by a group of dusty asymptotic giant branch stars. Conversely, this total infrared luminosity is too low and the temperatures are too uniform for the IR emission to be associated with extremely massive, evolved dust-enshrouded sources, such as Wolf-Rayet stars ($L_* > 10^5$ $L_\odot$). We therefore rule out the possibility that a local heating source is responsible for the observed dust temperatures.

Dust heating due to collisions with electrons is commonly observed in SNRs given the energetic conditions of the ejecta (e.g. *29*). Given the apparent location of the dust outside the regions of hard X-ray emission (see Fig. 1B), the dust will be in a cooler and denser region of the ejecta. Taking the electron density and temperature to be $\sim 100$ cm$^{-3}$ and $\sim 10^5$ K, which is consistent with conditions that result in a short ejecta cooling timescale ($\tau_{cool} << 10^4$ yr), we find that the radiation by the central cluster will dominate the heating over electron collisions for all grain sizes.

In the final scenario, we propose that the grains composing the proposed SNR dust are smaller than the typical $\sim 0.1$ μm size (*24*) and are therefore heated to higher temperatures due to lower heat capacities. The 31/37 dust temperature map in Fig. 3 is overlaid with several contours of the predicted dust equilibrium temperatures for 0.1 μm silicate and 0.01 μm amorphous carbon grains, where we have only considered radiative heating from the central cluster and assumed no blockage by the CND. The 0.1 μm dust temperature contours show clear agreement with the observed temperatures in the CND as well as the Northern Dust Cloud; however, the contours underpredict the temperatures observed for the proposed SNR dust. Shifting to the smaller 0.01 μm-sized grains shows much closer agreement between the predicted and observed dust temperatures ($\sim 100$ K).

We conclude that the inconsistency between the observed dust temperature of the proposed SNR dust and the Northern Dust Cloud and CND is due to a composition of smaller grains, which is consistent with expectations for dust in a SNR where fragmentation and thermal sputtering have taken place following grain formation.

**Dust Spectral Energy Distribution Models.** The observed spectral energy distribution (SED) provides important constraints on the size of the dust particles, their physical location and the heating source(s). With this in mind, we investigate six different regions in and around the SNR dust (see Fig. 4) and utilize the DustEM code (*30*) to produce dust models that fit the observed SEDs. Three of the regions cover sites of prominent far-IR emission: north clump (NC), south clump (SC), and east (E). The remaining three regions cover sites of strong emission observed at mid-IR and sub-mm wavelengths that may or may not be associated with the SNR: southeast (SE), west (W), and north (N). For the models, we perform a linear least-squares fit to the dereddened fluxes observed at 5.8, 8.0, 19.7, 25.2, 31.5, and 37.1 μm and assume the dust is composed of two independent grain distributions: $\sim 0.04$ μm-sized (large grain; LG) and small $\sim 0.001$ μm-sized (very small grain; VSG) amorphous carbon grains heated radiatively by the central cluster which we modeled as a point source having a luminosity of $4 \times 10^7$ $L_\odot$ (*31*) and the spectrum of a Castelli & Kurucz (2004) (*32*) stellar atmosphere with an effective temperature of 37000 K, representative of the luminous stars that dominate the radiation from the cluster. Because the distance between the central cluster and the proposed SNR dust, $d$, is uncertain, we allow $d$ to vary as a free parameter in our models. The best-fit models (Fig. 4, Table 1) show that $d$ is consistent with the ~ 5 pc separation distance between the SNR and Sgr A* estimated by kinematics of maser spots associated with the SNR (*11*).

The LG and VSG mass abundances are the other two free parameters for the models in addition to the distance between the central cluster and the dust. We note that assuming a uniquely larger-sized (0.1 μm) dust distribution results in very poor SED fits for all of the regions except W. We therefore require an independent distribution of VSGs for which transient heating allows us to fit the 5.8 and 8.0 μm flux points. A modified blackbody fit to the 5.8 and 8.0 μm points yields temperatures of ~ 350 K, which cannot be achieved if the dust is equilibrium-heated. The results from our best-fit models show that the VSG to LG mass abundance ratio for all the regions except W ranges from 15 - 90% (Fig. 4), which suggests a VSG enhancement when compared to the typical ratio of the Milky Way ISM (~ 13%; *33*) as well as that of the compact Sgr A East HII regions in the 50 km/s cloud ~ 3 pc east of the proposed SNR dust (~ 2 – 4 %; *25*).

To produce adequate fits to the SED of region W, we are required to either increase $d$ to ~ 10 pc or to increase the larger grain size to 0.2 μm. Since $d$ ~ 10 pc would suggest that the region W dust is located a factor of ~ 2 times further away than the dust at the other regions, we adopt the more likely interpretation that the dust grains are larger. Regardless of interpretation, the resulting inconsistent model fits to the region W SED strongly suggest that region W is not associated with the other regions, and might thus be an extension of the dust distribution of the CND, as its proximity to the CND might suggest.

Our model fits do not constrain the composition of the LG distribution, which can also be modeled with silicate grains. Substituting amorphous carbon for silicates in the models decreases the predicted total dust mass by only ~ 10%. Silicates for the VSG distribution, however, are ruled out as they yield poor fits due to the steepness of the shorter wavelength side of the 9.7 μm feature. Altering the size of the LG distribution by +/- 0.02 μm only changes the total mass by ~20%. PAHs are also ruled out for the VSG composition based on the ratio of the 7.7 μm and 8 μm fluxes (*23*). The resulting SED dust models and the fitted parameters are shown in Fig. 4 and Tab. 1, respectively.

**Sub-mm Excess and Dust Mass.** In our model fitting, we omit the 70 μm flux in order to determine whether the 70 μm flux is associated with the warm dust emission probed by the far-IR or is enhanced relative to the model due to the presence of a cooler dust distribution. Owing to the high background fluctuations at 70 μm images throughout the Galactic center, we assume the extracted 70 μm fluxes shown in the SEDs (Fig. 4) are upper limits. The absence of a 70 μm excess as well as the lack of ionized gas emission (e.g. Paschen-α; *27, 28*) spatially coincident with the proposed SNR dust indicates that it is unlikely to be associated with a molecular cloud along the line of sight towards the SNR.

The dust model fit to the SED of the full region covering the proposed SNR dust yields a total dust mass of ~ $0.02^{+0.008}_{-0.006}$ $M_\odot$, where the LG component composes ~ 45% of the total mass. This mass estimate is consistent with the summed dust mass of the 5 sampled regions excluding W (~ 0.013 $M_\odot$) as they cover approximately ¾ of the angular size of the SNR dust emission.

**Sgr A East Expansion and Dust Survival in a Non-Uniform Density Medium.** The location, thermal structure, and SED of the dust indicate that it is very likely associated with the Sgr A East SNR. We describe a theoretical framework of the evolution of the SN ejecta to explain the morphology, size composition, and apparent location of the surviving dust within the SNR. Our interpretation is based on comparing the ejecta cooling timescales to the dust destruction timescales due to thermal sputtering during the expansion of the ejecta into an asymmetric surrounding medium.

Molecular line observations towards Sgr A East reveal the presence of dense molecular clouds that appear to be interacting with the SNR, especially at the northern and eastern edges of the remnant (*9, 34*). Newly condensed dust from the initial explosion is assumed to be present uniformly throughout the SN ejecta, and that the density of the surrounding medium to the NE and SW is $\sim 10^3$ cm$^{-3}$ (*14*) and $\sim 100$ cm$^{-3}$, respectively. This ambient density gradient implies that there will be an asymmetry in the ejecta densities as well as in the strength and speed of the forward shock and reverse shock, which will influence the survival/destruction timescales of dust within the ejecta. After passage of the reverse shock, the density of the ejecta to the NE will therefore be $\sim 10$ times greater than the ejecta to the SW, and the shocked gas temperatures to the NE will be $\sim 5$ times lower than that to the SW. Hard X-ray observations of the Sgr A East ejecta reveal an electron temperature of $T_e \sim 2 \times 10^7$ K and density of $n_e \sim 10$ cm$^{-3}$ (*14, 35*). This hot ejecta can be associated with the diffuse SW expansion (see Fig. 1B), whereas the NE ejecta, which is located at the projected position of the observed dust, will be unobservable in hard X-ray emission given initial temperature of $T_e \sim 4 \times 10^6$ K and density of $n_e \sim 100$ cm$^{-3}$.

The timescale for complete destruction due to thermal sputtering for $a = 0.04$ µm-sized grains is $\sim 3000$ yr in the hot, diffuse region of the ejecta and $\sim 1000$ yr in the cooler, dense region (*23*). These timescales do not include the erosion effects due to the kinetic sputtering that occurs when the SN-condensed initially encounters the reverse shock (*3*), which include calculations beyond the scope of this paper; therefore, our estimates apply to dust that has survived the initial kinetic sputtering and can be treated as an upper limit on the dust lifetimes. Given the estimated age of the SNR, this destruction timescale implies that a large fraction of the LGs ($a \sim 0.04$ µm) and VSGs ($a \sim 0.001$ µm) we adopt for our SED models will be destroyed. However, unlike the diffuse, hot SW ejecta, the dense, cooler ejecta at the NE undergoes significant radiative cooling that occurs on timescales much shorter than the dust sputtering lifetimes.

In the metal-enriched environment of SN ejecta, iron will dominate the radiative cooling at temperatures greater than $10^6$ K (*36*). Assuming there is $\sim 0.15$ M$_\odot$ of iron within the ejecta (*35*) with a factor of $\sim 10$ density enhancement in the NE with respect to that of the SW and that the ejecta is in collisional ionization equilibrium in each region, we estimate the cooling timescale of the NE ejecta to be $\tau_{cool} \sim 400$ yr, which is much shorter than the estimated sputtering timescale. Assuming the NE ejecta cools to as low as $T_e \sim 10^5$ K, the destruction timescales are $\sim 10^6$ yr for $a = 0.04$ µm-sized grains, which is sufficiently long – even for VSGs – to survive within the ejecta. As the ejecta cools to even lower temperatures, the thermal sputtering becomes negligible. Toward the SW, we find that the cooling timescale for the ejecta is $\tau_{cool} \sim 3 \times 10^4$ yr, consistent with the detection of hard X-rays there, given the estimated age of the SNR. Unfortunately, the significant extinction towards Sgr A East does not currently allow for observations of optical cooling lines or soft X-ray emission to directly confirm our hypothesis. However, the prominent spatial anti-correlation of the dust emission and the hard X-ray emission in Sgr A East strongly suggests that dust survival conditions are much more favorable in cooler regions of the ejecta (see Fig. 1B).

The SN-condensed dust is then injected into the surrounding ISM with minimal erosion from thermal sputtering as well as non-thermal kinetic sputtering. Assuming the progenitor of Sgr A East was co-moving with the surrounding medium, the relative velocity between the dust and the ISM can be estimated to be ~ 100 km s$^{-1}$ based on the age of the SNR (~ $10^4$ yr) and the approximate distance travelled by the SN-condensed dust from the apparent center of the SNR (~ 1 pc). After the dust has been slowed from 100 km s$^{-1}$ to velocities of 10 km s$^{-1}$ by collisions with the surrounding ISM, the decrease in grain radius due to kinetic sputtering will be less than 10% (*3, 37*).

**Grain-grain Collisions and the VSG-LG Mass Ratio.** The temperatures of the NE ejecta fall in a regime where grain-grain collisions become significant, if we assume that the thermal velocities of the dust grains are closely coupled to those of the gas. Unlike sputtering by ions, which erodes the grain atom by atom, grain-grain collisions are efficient at redistributing mass from larger ($a \sim 0.1$ μm) to smaller (a ≲ 0.005 μm) grains by shattering or fragmentation (*38*). We attribute the enhanced VSG-LG mass ratio of the SNR dust derived from the SED models (Fig. 4 and Tab. 1) to grain-grain collisions, which occur between 0.005 μm and 0.1 μm grains on timescales of ~ 60 yr (*23*). Grain-grain collisions and the fragmentation of larger into smaller grains therefore occur on timescales shorter than the age of the SNR and the dust sputtering lifetimes, which should lead to a significant enhancement in the small grain mass abundance relative to that of the interstellar medium, consistent with our models.

**Supernova Dust Mass Survival Fraction and Implications for Galaxies of the Early Universe.** Our results show that given a dense surrounding environment, SN-condensed dust can indeed survive the destructive passage of the reverse shock to be injected into the ISM. The mass of the Sgr A East SNR dust provides an estimate of the mass survival fraction of dust initially condensed in the ejecta. SNe dust production models for a progenitor with a mass of 13 - 20 $M_\odot$ predict that ~ 0.3 $M_\odot$ of dust forms in the ejecta (*2, 3*); therefore, our derived SNR dust mass of ~ 0.02 $M_\odot$ implies that 7% of the total initial dust mass survived the passage of the reverse shock. This number is very uncertain, however, because estimates of dust masses produced in SNe are not well constrained given the uncertainties in the microphysics that dictate grain formation.

Excluding SN1987A, infrared observations of SNRs much younger than Sgr A East reveal that far smaller quantities of dust are formed in the ejecta than predicted by models (≲ 0.1 $M_\odot$; *39* and references therein). If we assume 0.1 $M_\odot$ of dust is initially formed in SNe, which is the quantity of dust detected in Cassiopeia A and the Crab Nebula (*39*), our results imply that 20% of the initial dust mass survives the reverse shock to be injected into the ISM.

We apply our results to the dust formation/survival rates in the ISM of the galaxies in the early Universe. In a scenario where the galaxy's star formation history undergoes a single short-duration and intense burst followed by a calm period having a much lower rate of star formation, a dust yield of ~ 0.15 $M_\odot$ per SN is required to produce the large inferred quantities of dust (*40*). Our current results suggest that it is difficult to produce the observed dust mass in such galaxies if only ~10 – 20% of the dust survives the reverse shock. However, since stars in early Universe galaxies form in significantly denser regions than those in local galaxies (*41* and references therein) the dust mass survival rate is likely greater than that which we infer. Additionally, if each SNe produced as much dust as observed in the ejecta of SN1987A (~ 0.5 M$_\odot$; *42, 43*), SNe could reasonably account for the dust production. These findings are consistent with SNe being a dominant dust production mechanism in galaxies of the early universe (*7*).

**Acknowledgements**. We would like to thank the rest of the FORCAST team, Matt Hankins, George Gull, Justin Schoenwald, and Chuck Henderson, the USRA Science and Mission Ops teams, and the entire SOFIA staff. Additionally, we would like to thank Eli Dwek and the anonymous referees for their insightful comments. Z. L. acknowledges support from the Recruitment Program of Global Youth Experts. This work is based on observations made with the NASA/DLR Stratospheric Observatory for Infrared Astronomy (SOFIA). SOFIA science mission operations are conducted jointly by the Universities Space Research Association, Inc. (USRA), under NASA contract NAS2-97001, and the Deutsches SOFIA Institut (DSI) under DLR contract 50 OK 0901. Financial support for FORCAST was provided by NASA through award 8500-98-014 issued by USRA. Data presented in this paper can be accessed from Database S1 (*23*).

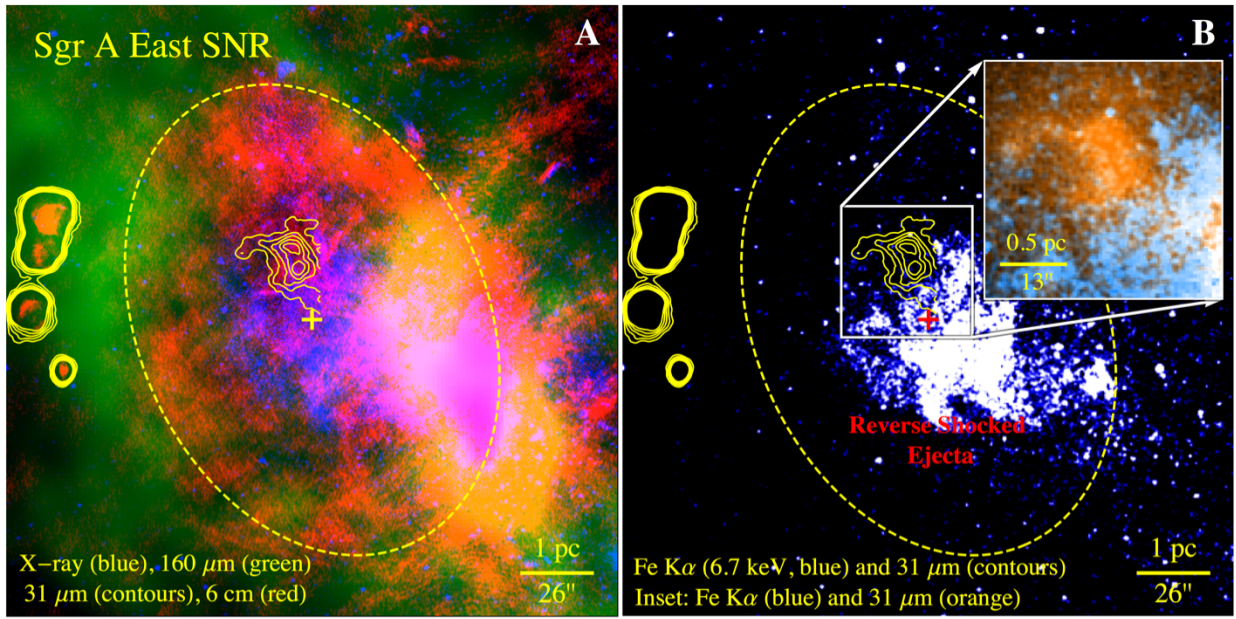

**Fig. 1.** (**A**) Composite false-color image of the Sgr A East SNR overlaid with contours of the 31.5 μm emission east of the Circumnuclear Disk (CND). North is up and east is left. The colors correspond to emission from 2 - 8 keV (blue; Chandra/ACIS-I), 160 μm (green; Herschel/PACS), and 6 cm (red; VLA). The dotted ellipse delineates the boundary of the Sgr A East radio shell with the cross indicating the location of its apparent center. (**B**) Fe Kα (6.7 keV; blue) emission from the SNR overlaid with the 31.5 μm emission contours and a 2 × zoomed inset of the SNR's infrared emitting region (orange).

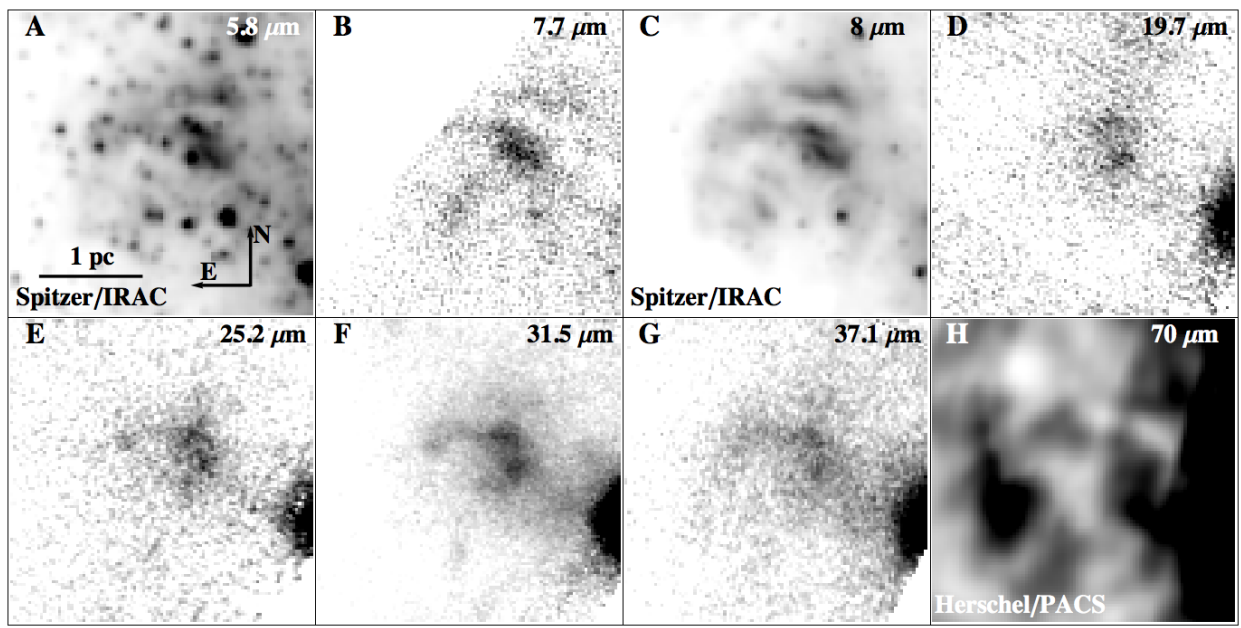

**Fig. 2.** (**A** to **H**) SOFIA/FORCAST, *Spitzer*/IRAC, and *Herschel*/PACS images of the Sgr A East SNR dust.

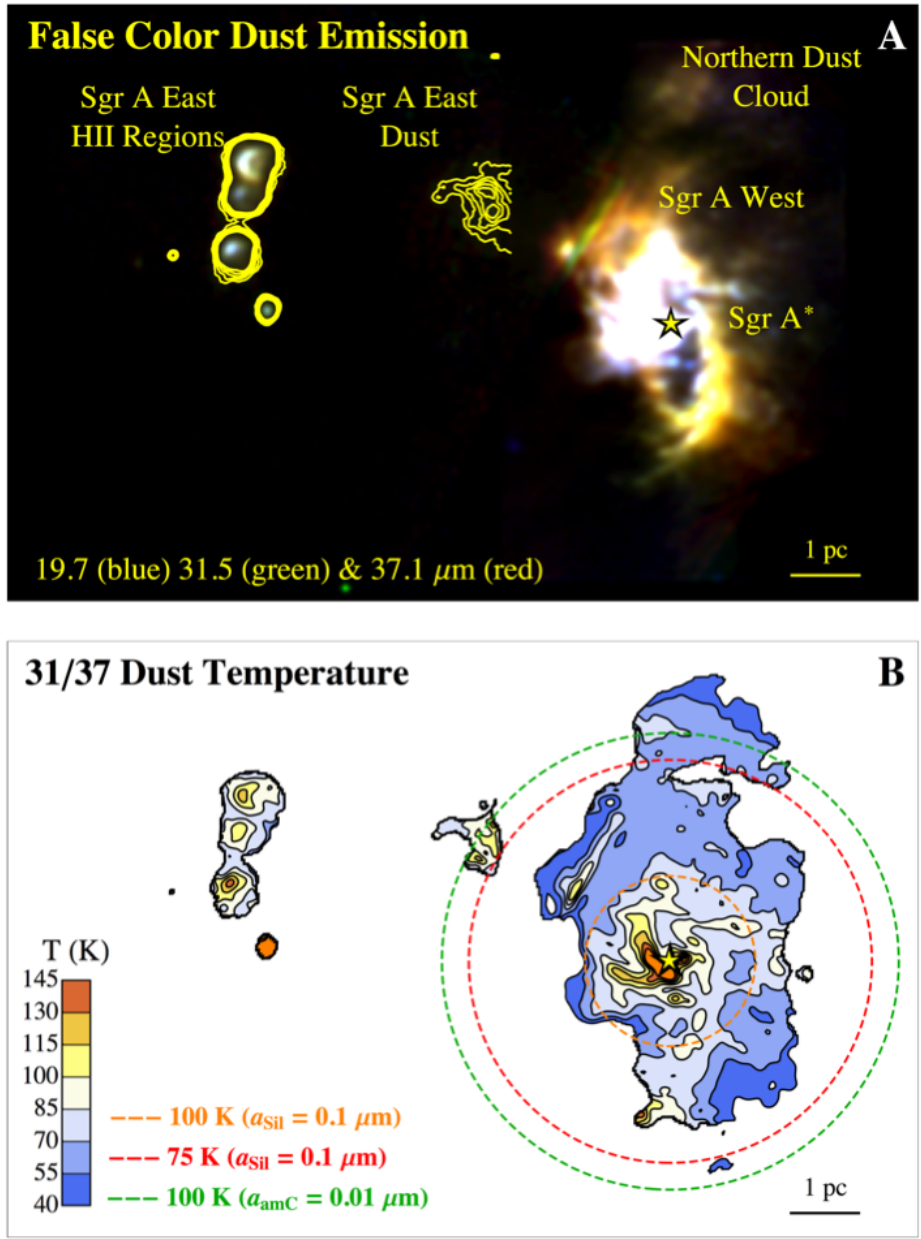

**Fig. 3.** (**A**) False color image of dust emission from the Sgr A West and East regions. The colors correspond to 19.7 (blue), 31.5 (green), and 37.1 (red) μm, and the 31.5 μm flux contours east of the CND are overlaid. The central cluster is located in the sub-parsec vicinity of Sgr A*. (**B**) 31/37 dust temperature map of the Sgr A West and East regions. Overlaid are the theoretical dust temperature contours at 75 and 100 K for 0.1 μm-sized silicate grains and at 100 K for 0.01 μm-sized amorphous carbon grains assuming equilibrium radiative heating by the central cluster. The apparent linear structure northeast of Sgr A* seen in the temperature contours is an artifact from combining the images of Sgr A East and Sgr A West, and the temperature peak at the south of Sgr A West is likely associated with an embedded source that appears more prominently in the mid-IR *Spitzer*/IRAC images (*18*). In both figures, north is up and east is left.

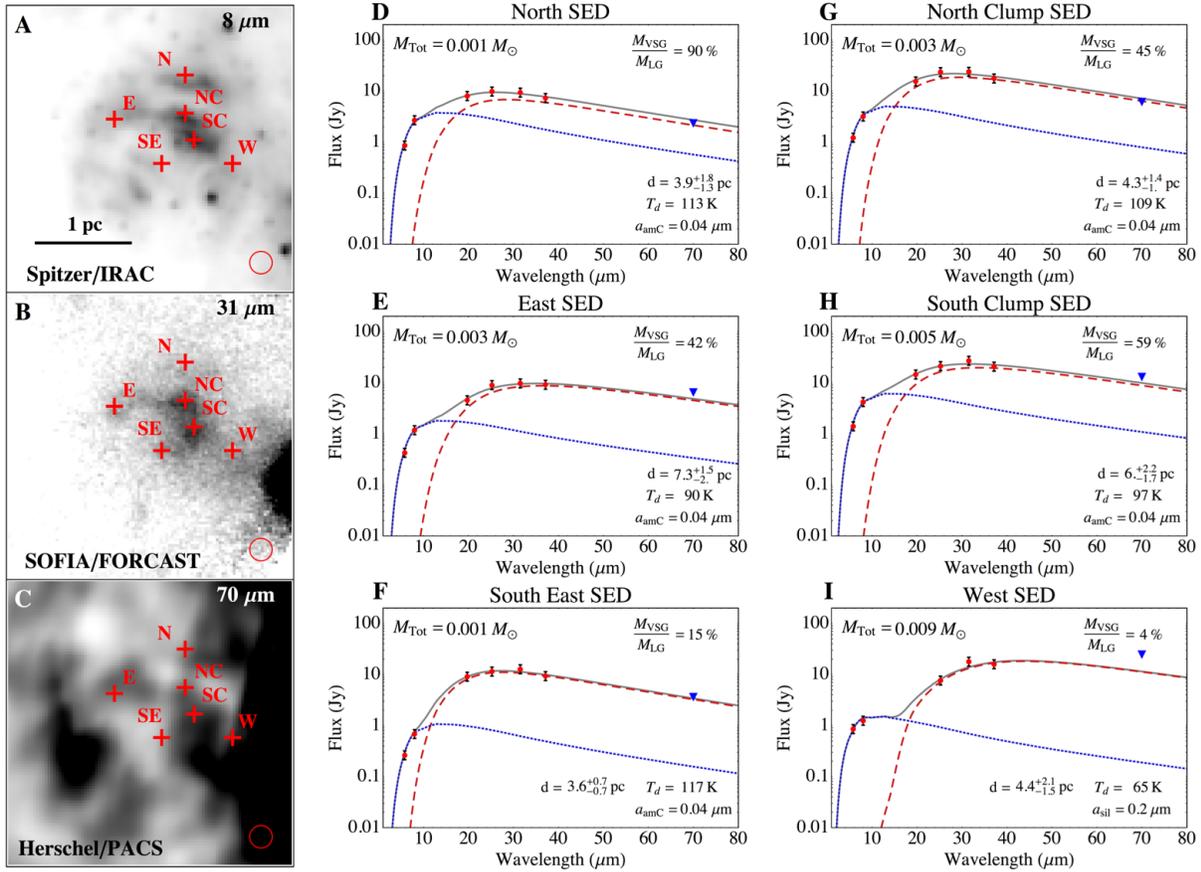

**Fig. 4.** (**A** – **C**) Sgr A East dust emission region. North is up and east is left. Overlaid are the six locations across the region where SEDs were extracted. The aperture size used to extract the fluxes is shown in the lower right corner. (**D** – **I**) Best-fit DustEM models of the six regions. The blue dotted line and the red dashed line correspond to emission from very small grain (VSG) and large grain (LG) distributions, respectively. The errors of both the *Spitzer*/IRAC and SOFIA/FORCAST fluxes are assumed to be 20%. Fluctuations in the amplitudes of the background emission in the 70 μm Herschel/PACS image can vary on the order of the source flux. Fluxes extracted from the 70 μm image are therefore treated as upper limits.

| Region | $a_{amC}^{VSG}$ | $a_{amC}^{LG}$ [a] | $T_d$ (K) | $d$ (pc) | $M_{Tot}$ ($\times 10^{-3}\,M_\odot$) | $M_{VSG}/M_{LG}$ |
|---|---|---|---|---|---|---|
| North | 0.001 | 0.04 | $113^{+19}_{-15}$ | $3.9^{+1.8}_{-1.3}$ | $1.1^{+1.1}_{-0.6}$ | $0.90^{+0.13}_{-0.14}$ |
| N. Clump | 0.001 | 0.04 | $109^{+11}_{-11}$ | $4.3^{+1.4}_{-1.0}$ | $2.6^{+1.9}_{-1.0}$ | $0.45^{+0.06}_{-0.06}$ |
| S. Clump | 0.001 | 0.04 | $97^{+12}_{-11}$ | $6.0^{+2.2}_{-1.7}$ | $5.3^{+4.7}_{-2.5}$ | $0.59^{+0.08}_{-0.09}$ |
| East | 0.001 | 0.04 | $90^{+8}_{-8}$ | $7.3^{+2.0}_{-1.5}$ | $2.9^{+1.9}_{-1.1}$ | $0.42^{+0.06}_{-0.06}$ |
| South East | 0.001 | 0.04 | $117^{+9}_{-8}$ | $3.6^{+0.7}_{-0.7}$ | $0.9^{+0.4}_{-0.3}$ | $0.15^{+0.02}_{-0.01}$ |
| West | 0.0008 | 0.2 | $65^{+10}_{-8}$ | $4.4^{+2.1}_{-1.5}$ | $9.2^{+14.3}_{-5.7}$ | $0.04^{+0.01}_{-0.01}$ |

**Table 1.** Assumed and derived dust properties of the dust SED model fit to the Sgr A East SNR dust. $a_{amC}^{VSG}$ and $a_{amC}^{LG}$ are the grain sizes assumed for the very small grain (VSG) and large grain (LG) component of the model given in units of micrometers. $T_d$ is the temperature of the LG distribution, $d$ is the distance between the dust in the region and the central stellar, $M_{Tot}$ is the total dust mass fit to the region, and $M_{VSG}/M_{LG}$ is the mass ratio of the VSGs to the LGs. Errors provided are 1-σ as determined from the weighted linear least-squares fits.

[a]The LG component for the model fit to the West region is composed of silicates, like the dust in the CND (*24*).

**Supplementary Materials:**

Materials and Methods

Supplementary Text

Figure S1

Table S1

External Database S1

References (*44 – 48*)

# Supplementary Materials for

## Old Supernova Dust Factory Revealed at the Galactic Center

R. M. Lau*, T. L. Herter, M. R. Morris, Z. Li, J. D. Adams

*Correspondence to: ryanl@astro.cornell.edu

**This PDF file includes:**

    Materials and Methods
    SupplementaryText
    Fig. S1
    Table S1

**Other Supplementary Materials for this manuscript includes the following:**

    Database S1

**Materials and Methods**

SOFIA/FORCAST Observations, Reduction, and Extinction Correction

Observations were made using FORCAST (*15*) on the 2.5-m telescope aboard SOFIA. FORCAST is a 256 × 256 pixel dual-channel, wide-field mid-infrared camera sensitive from 5 – 40 μm with a plate scale of 0.768" per pixel and field of view of 3.4' × 3.2'. SOFIA/FORCAST observations of the Sgr A East SNR were made on the Cycle 1 Flights 110 and 131 on July 2, 2013 and September 17, 2013, respectively, at an altitude of ~ 39,000 ft. Flight 110 observations were taken at 7.7 μm and Flight 131 observations were taken at 19.7, 25.2, 31.5, and 37.1 μm. Total on-source integration time was ~ 340 sec at 7.7 μm, ~ 100 sec at 19.7 and 25.2 μm, ~ 200 sec at 31.5 μm, and ~ 50 sec at 37.1 μm. The quality of the images was consistent with near-diffraction-limited imaging at the longest wavelength; the full width at half maximum (FWHM) of the point spread function (PSF) was ~ 3.2" at 19.7 μm and ~ 3.8" at 37.1 μm, in comparison to 5.2" and 12" of Herschel/PACS at 70 and 160 μm, respectively.

Calibration of the images was performed by observing standard stars and applying the resulting calibration factors as described in Herter *et al.* (2013) (*16*). Color correction factors were negligible (≲5%) and were therefore not applied. The 1-σ uncertainty in calibration due to photometric error, variation in water vapor overburden, and airmass is ±7%; however, due to flat field variations (~ 15%), which we are unable to correct for, we conservatively adopt a 1σ uncertainty of ±20%.

Large column densities of dust and gas lead to extreme extinction along lines of sight towards the Galactic center ($A_V$ ~ 30; *44*). We adopt the extinction curve derived by Fritz *et al.* (2011) (*44*) from hydrogen recombination line observations of the minispiral, the HII region in the inner 3 pc of the Galactic center, at 1 - 19 μm made by the Short Wave Spectrometer (SWS) on the Infrared Space Observatory (ISO) and the Spectrograph for Integral Field Observations in the Near Infrared (SINFONI) on the Very Large Telescope (VLT).

*Chandra*/ACIS-I Observations and Reduction

To obtain a high-quality X-ray image of Sgr A East, we utilized 48 *Chandra*/ACIS-I observations with a total effective exposure of 1.47 Ms. The raw data were downloaded from the *Chandra* public archive and reprocessed with the latest version of CIAO (v4.6) and the corresponding calibration files. The relative astrometry among individual observations was calibrated by matching centroids of bright, point-like sources detected within the common field-of-view. For each observation, we produced count and exposure maps in selected energy bands; corresponding instrumental background maps were generated from the ``stowed'' data, after calibrating with the 10-12 keV count rate. These individual maps were then reprojected to a common tangential point, here the position of Sgr A*, to produce a combined flux image. To produce a ``net'' flux image for the He-like Fe 6.7 keV line, we further contrasted an ``on-line'' (6.3-7.1 keV) image with an ``off-line continuum'' (6.0-6.3 keV plus 7.1-8.0 keV) image.

Probing Dust Composition with FORCAST 7.7 μm and IRAC 8 μm Observations

The different bandwidths of the FORCAST 7.7 μm (Δλ = 0.47 μm) and IRAC 8.0 μm (Δλ = 2.8 μm) filters provides us with a unique observational technique that allows us to probe the composition of the dust emitting at ~ 8 μm, which typically arises from stochastically heated, very small grains ($a \sim 0.001$ μm). Given the spectral profiles of different dust species at the mid-IR and the filter/instrument responses we can predict the relative flux difference measured between each instrument.

As shown in Fig. S1, FORCAST will measure a greater flux than IRAC if the emission is dominated by PAHs due to their prominent 7.7 μm feature. If the emission is dominated by very small silicates, the broad 9.7 μm feature will contribute significantly to the flux detected by IRAC which will be much greater than the FORCAST flux. The flatness of the spectra in the mid-IR from very small amorphous carbon implies that IRAC and FORCAST will measure similar fluxes. If the dust grains are large (i.e. not transiently heated) and cooler than ~ 100 K the flux measured by IRAC will be larger than the flux measured by FORCAST for amorphous carbon as well as silicates.

Applying the derredening prescription described by Fritz *et al.* (2011) (*44*) for filters of varying bandwidth, we find that the FORCAST 7.7 μm and IRAC 8.0 μm fluxes agree very closely with each other for the SNR dust (see Tab. S1), which suggests that the dust emitting at the mid-IR is dominated by hot amorphous carbon. This is consistent with the selection of amorphous carbon grains for the VSG component of our dust SED models. The lack of dominant PAH emission provides further evidence that this dust is indeed associated with the SNR since PAHs are not detected in SNRs.

**Supplementary Text**

<u>Collisional</u> <u>vs.</u> <u>Radiative</u> <u>Heating</u>

Collisional heating by electrons is typically a dominant heating mechanism for dust in shocked SNR ejecta; however, the proximity of Sgr A East to the luminous central stellar cluster suggests that the radiative heating contribution will be significant. To study the relative contributions from collisional and radiative heating we derive the ratio shown in Eq. S1 and apply it to different ejecta conditions,

$$\frac{\left(\frac{dE}{dt}\right)_{rad}}{\left(\frac{dE}{dt}\right)_{coll}} = \frac{F_* \langle Q \rangle_*(a)}{1.71 \times 10^{-10} n_e T_e^{3/2} h(a, T_e)}, \quad (S1)$$

where $a$ is the radius of the target dust grain, $\langle Q \rangle_*$ is the dust emission efficiency averaged over the incident radiation field, $F_*$ is the radiative flux from the heating source, and $h$ is the effective grain heating efficiency for electron-grain collisions (*45*). It is assumed that the dust is located in a dense, cooler region of the SN ejecta and heated radiatively by the central cluster, which has a luminosity of ~ $4 \times 10^7$ L$_\odot$ (*31*). We adopt a separation distance between the central cluster and the SNR dust of its projected distance (~ 3 pc) multiplied by a factor of √2. Although higher electron densities implies a higher dust-electron collision rate, the electron temperatures will be lower due to the attenuation of the shock strength in the denser medium as well as enhanced radiative cooling. Radiation from the central cluster therefore dominates the heating of the dust in dense regions of the ejecta:

$$\left(\frac{dE}{dt}\right)_{rad} > \left(\frac{dE}{dt}\right)_{coll} \text{ for } T_e \sim 10^5 \text{ K and } n_e \sim 100 \text{ cm}^{-3}. \tag{S3}$$

We note that the UV emission from the cooling of the metals composing the cool, dense ejecta in the vicinity of the SN-condensed dust ($T_e \sim 10^5$ K and $n_e \sim 100$ cm$^{-3}$) may contribute significantly to dust heating relative to the radiative heating from the central cluster:

$$\frac{\left(\frac{dE}{dt}\right)_{rad,*}}{\left(\frac{dE}{dt}\right)_{rad,ej}} \approx \frac{F_*}{\Lambda_i(T_e) n_e n_i l}, \tag{S4}$$

where $\Lambda_i$ is the atomic cooling function of the ionic species $i$ with density $n_i$ and $l$ is the characteristic length scale of the ejecta. Assuming that oxygen is the dominant cooling element in the ejecta at a temperature of $\sim 10^5$ K (*36*) and that the oxygen number density is $n_O \sim n_e/8$ (see below), $\Lambda_O$ is $\sim 10^{-19}$ ergs s$^{-1}$ cm$^3$ (*36*) and Eq. S4 evaluates to a value of order unity. However, given that the dense ejecta region temperatures are uncertain and that not all of the ejecta cooling goes into radiation that is absorbed by the dust since cooling will occur through other processes such as adiabatic expansion and free-free emission, a more rigorous calculation of each cooling mechanism and a measurement of the ejecta temperature is required to determine the relative importance of the radiative cooling from the ejecta in heating the dust. Ultimately, if the radiation from cooling is found to be a significant contributor to heating the dust, it does not affect our results nor does it require any re-interpretation of the SN-condensed dust; it will, in fact, strengthen our interpretation that the dust is associated with the SNR and present in a cool, dense region of the ejecta.

Dust Sputtering and Shattering

In order to study the survival/destruction of SNR dust it is particularly informative to determine sputtering timescales given the observed properties of the hot gas composing the ejecta. The lifetime of a dust grain of size, $a$, in a hot gas with thermal velocity, $v$, can be estimated by the sputtering rate,

$$\frac{da}{dt} = \frac{m_a}{2\rho_b} n_i \langle Y_i(v) v \rangle \tag{S5}$$

where $m_a$ is the mass of the sputtered atoms, $\rho_b$ is the bulk density of the grain species, and $n_i$ and $\langle Y_i(v) v \rangle$ are the density and velocity averaged sputtering yield of the gas species $i$. Since observations of the ejecta only yield measurements of the electron density and temperature and not those of the sputtering ions, it is difficult to accurately derive the sputtering rate. Similar to the dust sputtering calculations in the ejecta of SNR 1E 0102.2-7219 performed by Sandstrom *et al.* (2009) (*46*), we assume that oxygen is the most abundant metal in the ejecta and dominates the thermal sputtering in the reverse shock and that $T_O \sim T_e$. We also assume that oxygen in the ejecta is entirely ionized and that $n_O \sim n_e/8$.

For $T_e \sim 2 \times 10^7$ K, $n_e \sim 10$ cm$^{-3}$ ($n_O \sim 1$ cm$^{-3}$), and sputtering yields adopted from Tielens *et al.* (1994) (*47*), the dust lifetime can be expressed as

$$\tau_{sp} \sim 3000 \left(\frac{a}{0.04\ \mu m}\right)\left(\frac{n_O}{1\ cm^{-3}}\right)^{-1} yr \tag{S6}$$

In denser regions of the ejecta, the plasma temperatures are lower due to the slowing of the reverse shock and the increased efficiency of radiative cooling. For the conditions in these dense regions before radiative cooling, $T_e \sim 4 \times 10^6$ K and $n_e \sim 100$ cm$^{-3}$ ($n_O \sim 10$ cm$^{-3}$), the dust lifetime is

$$\tau_{sp} \sim 1000 \left(\frac{a}{0.04\ \mu m}\right)\left(\frac{n_O}{10\ cm^{-3}}\right)^{-1} yr \tag{S7}$$

After radiative cooling, $T_e \sim 10^5$ K and $n_e \sim 100$ cm$^{-3}$ ($n_O \sim 10$ cm$^{-3}$), the dust lifetime is much longer due to the lower temperatures despite an increase in the density,

$$\tau_{sp} \sim 10^6 \left(\frac{a}{0.04\ \mu m}\right)\left(\frac{n_O}{10\ cm^{-3}}\right) yr \tag{S8}$$

Dust is observed in dense, radiatively shocked regions of the ejecta in other SNRs such as Cassiopeia A (*48*) and 1E 0102.2-7219 (*46*). The velocity of the reverse shock in these dense regions is slowed below the complete dust destruction threshold from sputtering and falls into the regime where grain-grain collisions become significant ($v \sim 50 - 200$ km/s).

Grain-grain collisions will occur most frequently between small ($a \lesssim 0.005$ μm) and large grains ($a \sim 0.1$ μm). For projectiles that are much smaller than the target grains ($a_{VSG} \ll a_{LG}$) the grain-grain collision timescale can be approximated as

$$\tau_{coll} \sim \frac{1}{n_{VSG}\ \pi\ a_{LG}^2\ v_{coll}} \tag{S9}$$

where $v_{coll}$ is the velocity of the collisions and $n_{VSG}$ is the density of the projectile VSGs. We estimate $n_{VSG}$ by assuming the VSGs compose a fraction, $\xi_{VSG}$, of the total initial dust mass in the ejecta, $M_d$, and of volume, $V_{ej}$:

$$n_{VSG} \sim \frac{M_d\ \xi_{VSG}}{4/3 \rho_b\ \pi\ a_{VSG}^2\ V_{ej}}. \tag{S10}$$

If the grain size distribution can be approximated as a power-law with an index similar to that of an MRN distribution, where $n(a) \propto a^{-3.5}$, then $\xi_{VSG} \sim a_{VSG} / a_{LG}$. For an ejecta radius of $r_{ej} \sim 0.5$ pc with a total dust mass of $M_d \sim 0.02\ M_\odot$, we can rewrite the grain-grain collision timescale as

$$\tau_{coll} \sim 60 \left(\frac{r_{ej}}{0.5\ pc}\right)^3 \left(\frac{a_{VSG}}{0.005\ \mu m}\right)^2 \left(\frac{a_{LG}}{0.1\ \mu m}\right)^2 \left(\frac{M_d}{0.02\ M_\odot}\right)^{-1} \left(\frac{v_{coll}}{50\ km/s}\right)^{-1} yr \tag{S11}$$

Each collision can shatter the target grain into smaller fragments ranging from $\sim 0.001$ μm to 1/5 the size of the target grain, where the mass from the target grain is predominantly redistributed to the smaller fragment sizes (*38*). Shattering is much less efficient for target grains below a size of ~ 0.04 μm for ~ 100 km/s velocity shocks (*38*), which is why we have adopted *a* = 0.04 μm as the LG size for our SED models.

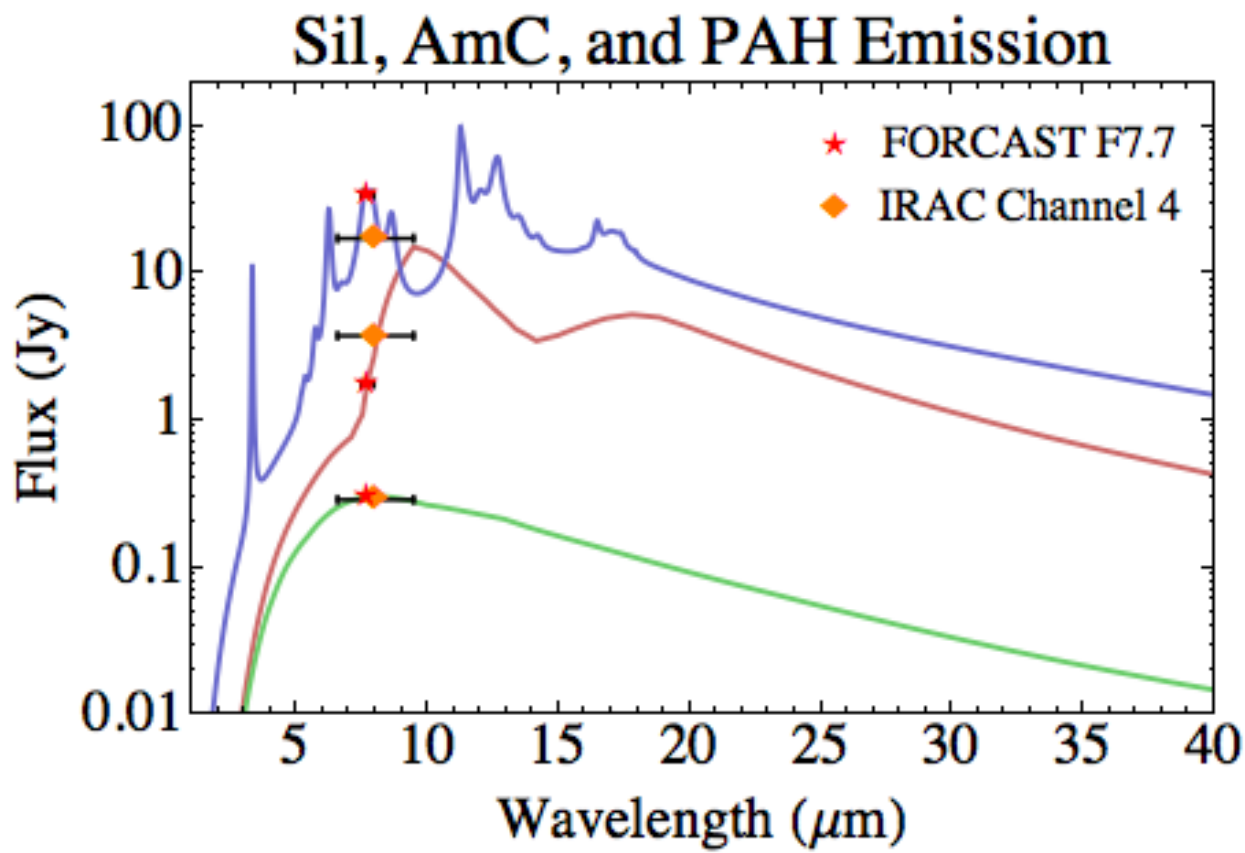

**Fig. S1.**
Estimated flux measurement of the model emission from three transiently-heated dust species. The error bars show the widths of the 7.7 μm FORCAST filter (red star) and the Channel 4 IRAC filter (orange diamond). Arbitrarily normalized, predicted fluxes are shown for three species of very small grains (< 0.001 μm): PAHs (blue), silicates (red), and amorphous carbon (green).

**Table S1.**

Observed IR dust emission from Sgr A East. A circular 5''-radius aperture was used to extract the flux (in Janskys). Flux from the dust near the center of the SNR is detected at a level of 4-σ above the background at 7.7 and 19.7 μm, 6-σ at 25.2 and 37.1 μm, and 15-σ at 31.5 μm. The 1-σ flux levels, which were determined statistically from an aperture placed over the background emission, are 0.006, 0.013, 0.024, 0.015, and 0.031 Jy per pixel at 7.7, 19.7, 25.2, 31.5, and 37.1 μm, respectively. Fluctuations in the amplitudes of the background emission in the 70 μm Herschel/PACS image can vary on the order of the source flux. Fluxes extracted from the 70 μm image are therefore treated as upper limits. Fluxes are not listed where the signal-to-noise ratio of the source is too low.

| Region | $F_{5.8}$ | $F_{7.7}$ | $F_{8.0}$ | $F_{19}$ | $F_{25}$ | $F_{31}$ | $F_{37}$ | $F_{70}$ |
|---|---|---|---|---|---|---|---|---|
| North | 0.37 | | 0.83 | 1.99 | 4.06 | 5.31 | 4.86 | 2.0 |
| N. Clump | 0.53 | 1.2 | 0.99 | 3.84 | 9.72 | 13.6 | 11.8 | 5.3 |
| S. Clump | 0.61 | 1.7 | 1.30 | 3.62 | 8.81 | 15.4 | 13.8 | 11.3 |
| East | 0.18 | | 0.37 | 1.13 | 3.73 | 5.54 | 6.10 | 5.5 |
| South East | 0.11 | | 0.21 | 2.26 | 4.75 | 7.23 | 6.22 | 3.2 |
| West | 0.37 | | 0.39 | | 3.16 | 10.2 | 10.6 | 21.5 |

**Additional Database S1 (Supplemental Data.zip).**

Mid-IR, far-IR, sub-mm, radio, and X-ray images utilized in our analysis of the Sgr A East SNR. 5.8 and 8 μm *Spitzer*/IRAC data are labeled SgrAE5_8.fits and SgrAE8.fits, respectively. SOFIA/FORCAST images at 7.7, 19.7, 25.2, 31.5, and 37.1 μm are labeled SgrAE7_7.fits, SgrAE19.fits, SgrAE25.fits, SgrAE31.fits, and SgrAE37.fits, respectively. *Herschel*/PACS images at 70 and 160 μm are labeled SgrAE70.fits and SgrAE160.fits, respectively. The VLA 6 cm radio image is labeled SgrAE6cm.fits. The *Chandra*/ACIS-I X-ray continuum and Fe Kα images are labeled SgrAEW_flux_2_8.fits and SgrAEW_flux_6.3_7.1.fits, respectively.